\documentclass[twoside,fleqn]{article}
\usepackage[headings]{thisespcrc2}
\usepackage{amsmath,graphicx,axodraw}

\allowdisplaybreaks
\newcommand{\eg}{e.g.\ }
\newcommand{\ie}{i.e.\ }

\newcommand\FA{\textit{FeynArts}}
\newcommand\FC{\textit{FormCalc}}

\newcommand\LT{\textit{LoopTools}}
\newcommand\cuba{\textsc{Cuba}}

\newcommand{\M}{\mathcal{M}}
\newcommand{\ri}{\mathrm{i}}

\newcommand{\braket}[2]{\left\langle #1\vphantom{#2}%
  \right. \kern-2.5pt\left| #2\vphantom{#1}\right\rangle}

\newcommand{\diracslash}[1]{\setbox0=\hbox{$#1$}%
  \rlap{\ifdim\wd0>.7em\kern.22\wd0\else\kern.1\wd0\fi /}#1}

\newcommand{\uscore}{\symbol{95}}

\begin{document}

\title{New Features in \FC\ 4}

\author{Thomas Hahn\address{%
	Max-Planck-Institut f\"ur Physik\hfill {\small MPP--2004--71} \\
        F\"ohringer Ring 6 \\
	D--80805 Munich, Germany}}

\begin{abstract}
\FC\ is a Mathematica package for the automatic computation of 
tree-level and one-loop Feynman amplitudes.  It accepts diagrams 
generated by \FA, simplifies them, and generates a complete Fortran code 
for their numerical evaluation.  Version 4 includes new features which 
enhance performance, convenience of use, and modularity/code 
reusability.
\end{abstract}

\maketitle


\section{Introduction}

\FC\ \cite{FormCalc} is a Mathematica package for the calculation of
Feynman amplitudes.  It prepares the amplitudes generated by \FA\
\cite{FeynArts} for numerical evaluation.  This consists of an algebraic
simplification step, including \eg the tensor reduction and the
calculation of fermionic traces, and a code generation step, where the
complete Fortran code for the evaluation of the squared matrix element
is written out.  Currently, diagrams up to one loop can be simplified,
and kinematics are supplied for $1\to 2$, $2\to 2$, and $2\to 3$
processes.

The present article describes the new features added in version 4.  They 
can be classified into

\smallskip
\noindent\emph{Performance features:} \\
-- Weyl--van der Waerden (WvdW) formalism, \\
-- phase-space integration by the \cuba\ library, \\
-- parallelization by shell script,

\smallskip
\noindent\emph{Convenience features:} \\
-- log-file management, \\
-- a simple way to resume aborted calculations, \\
-- shell scripts to perform common tasks,

\smallskip
\noindent\emph{Modularity/reusability features:} \\
-- general-purpose utilities library, \\
-- organization into master- and sub-makefiles, \\
-- symbol prefixing to avoid name conflicts.

\smallskip
\noindent Furthermore, the new \texttt{FeynInstall} script greatly 
simplifies the installation or upgrade of the \FA, \FC, and \LT\ 
packages.


\section{Weyl--van der Waerden formalism}

Amplitudes involving external fermions have the form
\begin{equation}
\M = \sum_{i = 1}^n c_i\, F_i\,,
\end{equation}
where the $F_i$ are (products of) fermion chains.  The textbook 
recipe is to compute probabilities, such as
\begin{equation}
|\M|^2 = \sum_{i, j = 1}^n c_i^*\, c_j\, F_i^* F_j\,,
\end{equation}
and evaluate the $F_i^* F_j$ by standard trace techniques.

The problem with this approach is that instead of $n$ of the $F_i$ one
needs to compute $n^2$ of the $F_i^* F_j$.  Since essentially $n\sim
(\text{number of vectors})!$, this quickly becomes a limiting factor in
problems involving many vectors, \eg in multi-particle final states or
polarization effects.

The solution is of course to compute the amplitude $\M$ directly and
this is done most conveniently in the WvdW formalism \cite{WvdW}.  The
implementation of this technique in an automated program has already
been outlined in \cite{Optimizations} and is now tested and available in
\FC\ 4.

The \texttt{FermionChains} option of \texttt{CalcFeynAmp} determines how
fermion chains are returned: \texttt{Weyl}, the default, selects Weyl
chains.  \texttt{Chiral} and \texttt{VA} select Dirac chains in the
chiral ($\omega_+/\omega_-$) and vector/axial-vector ($1/\gamma_5$)
decomposition, respectively.  The Weyl chains do not need to be further
evaluated with \texttt{HelicityME}, which applies the trace technique.

The WvdW method has other advantages, too: Polarization does not `cost'
extra in terms of CPU time, that is, one gets the spin physics for free. 
Whereas with the trace technique the formulas become significantly more
bloated when polarization is taken into account, in the WvdW formalism
one actually needs to sum up the polarized amplitudes to get the
unpolarized result.

There is also better numerical stability because components of $k^\mu$ 
are arranged as `large' and `small' matrix entries, viz.
\begin{equation}
\sigma_\mu k^\mu = \begin{pmatrix}
k_0 + k_3 & k_1 - \ri k_2 \\
k_1 + \ri k_2 & k_0 - k_3
\end{pmatrix}.
\end{equation}
Cancellations of the form $k_0 - k_3 = \sqrt{k^2 + m^2} - \sqrt{k^2}$ 
for $m\ll k$ are avoided and hence mass effects are treated more 
accurately.


\section{Phase-space integration}

The recently completed \cuba\ library \cite{Cuba} has been integrated
into \FC\ 4.  It provides four subroutines for multidimensional
numerical integration.  All four have a very similar invocation and can
thus be interchanged easily, \eg for comparison.  The flexibility of a
general-purpose method is particularly useful in the setting of
automatically generated code.

The following algorithms are included:

\textit{Vegas} is the classic Monte Carlo algorithm which uses
importance sampling for variance reduction.  It iteratively builds up a
piecewise constant weight function, represented on a rectangular grid. 
Each iteration consists of a sampling step followed by a refinement of
the grid.  The present implementation uses Sobol quasi-random numbers
for sampling.

\textit{Suave} is a crossover between Vegas and Miser and combines
Vegas-style importance sampling with globally adaptive subdivision:
Until the requested accuracy is reached, the region with the largest
error is bisected along the axis in which the fluctuations of the
integrand are reduced most.  In each half the number of new samples is 
prorated for the fluctuation.

\textit{Divonne} is a further development of the \textsc{CernLib}
routine D151.  It is intrinsically a Monte Carlo algorithm but has
cubature rules built in for comparison, too.  The variance-reduction
method is stratified sampling.  In a first step, a tessellation of the
integration region is constructed in which all subregions have an
approximately equal value of the spread, defined as
\begin{equation}
s(r) = \frac{\mathop{\mathrm{Vol}}(r)}{2}
   \Bigl(\max_{\vec x\in r} f(\vec x\,) -
         \min_{\vec x\in r} f(\vec x\,)\Bigr).
\end{equation}
Minimum and maximum here are sought using methods from numerical
optimization.  The subregions are then sampled independently with a
number of points extrapolated to reach the required accuracy.  For each
region, the latterly obtained value is compared to the initial rough
estimate and if the two are not compatible within their errors, the
region is subdivided or sampled once more.  Additions with respect to
the \textsc{CernLib} version are the final comparison phase and the
possibility to point out known extrema, to speed up convergence.

\textit{Cuhre} is a new implementation of \textsc{dcuhre}.  It is a
deterministic algorithm which employs cubature rules of a polynomial
degree.  Variance reduction is by globally adaptive subdivision: Until
the requested accuracy is reached, bisect the region with the largest
error along the axis with the largest fourth difference.

\begin{figure}
\centerline{\includegraphics[width=.9\linewidth]{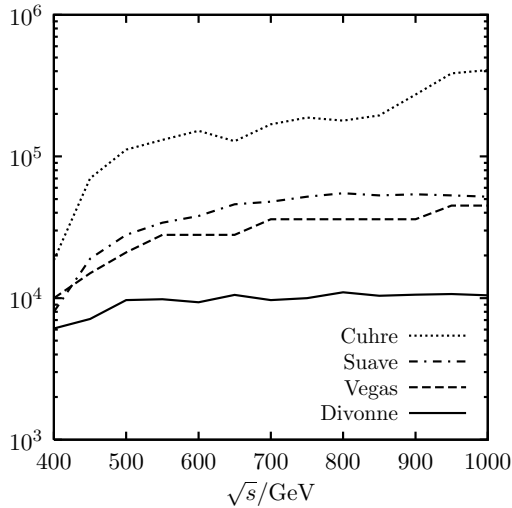}}
\caption{\label{fig:eettA}The number of integrand evaluations for 
the phase-space integration of $e^+e^-\to\bar t t\gamma$ at a requested 
relative accuracy of $10^{-3}$.}
\end{figure}

Fig.\ \ref{fig:eettA} compares the performance of the four algorithms
for a real phase-space integration of the process $e^+e^-\to\bar t
t\gamma$.  Above all it is very important to have several independent
integration methods to cross-check the results.

\section{Parallelization}

Calculations in models like the MSSM, where not all input parameters are
yet known, often require extensive scans to cover an interesting part of
the parameter space.  Such a scan can be a real CPU hog, but on the
other hand, the calculation can be performed completely independently
for each parameter set and is thus an ideal candidate for
parallelization.  The real question is thus not how to parallelize the
calculation, but how to automate the parallelization.

In \FC, the user may specify parameter loops by defining preprocessor 
variables, \eg
\begin{verbatim}
#define LOOP1 do 1 TB = 2, 30
\end{verbatim}
These definitions are substituted at compile time into a main loop of 
the form \\[1ex]
\verb|      LOOP1| \\
\verb|      LOOP2| \\
\verb|        |$\vdots$ \\
\verb|      |\textit{calculate cross-section} \\
\verb|  1   continue| \\[1ex]
The obstacle to automatic parallelization is that the loops are 
user-defined and in general nested.  A serial number is introduced to 
unroll the loops: \\[1ex]
\verb|      s = 0| \\
\verb|      LOOP1| \\
\verb|      LOOP2| \\
\verb|        |$\vdots$ \\
\verb|      s = s + 1| \\
\verb|      if( s |\textit{not in allowed range}\verb| ) goto 1| \\
\verb|      |\textit{calculate cross-section} \\
\verb|  1   continue| \\[1ex]
The serial number range can be specified on the command line so that it
is quite straightforward to distribute patches of serial numbers on
different machines.  Most easily this is done in an interleaved manner,
since one then does not need to know to which upper limit the serial
number runs, \ie if there are $N$ machines available, send serial
numbers 1, $N + 1$, $2N + 1$, etc.\ on machine 1, send serial numbers 2,
$N + 2$, $2N + 2$, etc.\ on machine 2, \dots

This procedure is completely automated in \FC\ 4: The user once creates
a \texttt{.submitrc} file in his home directory and lists there all
machines that may be used, one on each line.  In the case of
multi-processor machines he puts the number of processors after the host
name.

The executable compiled from \FC\ code, typically called \texttt{run}, 
is then simply prefixed with \texttt{submit}.  For instance, instead of
\begin{verbatim}
   run uuuu 500,1000
\end{verbatim}
the user invokes
\begin{verbatim}
   submit run uuuu 500,1000
\end{verbatim}
The \texttt{submit} script uses \texttt{ruptime} to determine the load 
of the machines and \texttt{ssh} to log in.  Handling of the serial 
number is invisible to the user.


\section{Log-file management and Resume}

Due to the parallelization mechanism, a single output file is no longer
sufficient.  Instead of a single log file, \FC\ 4 creates a log
directory, opens one log file for each serial number in this directory,
and redirects console output to this file.

Each log file contains both the `real' data and the `chatter' (progress,
warning, and error messages).  This has the advantage that no unit
numbers must be passed between subroutines -- every bit of output is
simply written to the console (unit \texttt{*} in Fortran).  It also
makes it easier to pinpoint errors, since the error message appears
right next to the corrupted data.  The `real' data are marked by an @ in
column 1 and there exists a simple shell script, \texttt{data}, to 
extract the real data from the log file.

The new log-file management also provides an easy way to resume an
aborted calculation.  This works as follows: when running through the
loops of a parameter scan, the log file for a particular serial number \\
-- may not exist: \\
\hphantom{--} then it is created with execute permissions, \\
-- may exist, but have execute permissions: \\
\hphantom{--} then it is overwritten, \\
-- may exist and have read-write permissions: \\
\hphantom{--} then this serial number is skipped. \\
The execute permissions, which serve here merely as a flag to indicate
an ongoing calculation, are reduced to ordinary read-write permissions
when the log file is closed.

In other words, the program skips over the parts of the calculation that
are already finished, so all the user has to do to resume an aborted 
calculation is start the program again with the same parameters.


\section{Shell scripts}

\FC\ 4 includes a few useful shell scripts:

\smallskip
\texttt{sfx} packs all source files (but not object, executable, or log
files) in the directory it is invoked in into a mail-safe
self-extracting archive.  For example, if \texttt{sfx} is invoked in the
directory \texttt{myprocess}, it produces \texttt{myprocess.sfx}.  This
file can \eg be mailed to a collaborator, who needs to say
``\texttt{myprocess.sfx x}'' to unpack the contents.

\smallskip

\texttt{pnuglot} produces a high-quality plot in Encapsulated PostScript
format from a data file in just one line.  In fact, \texttt{pnuglot}
does not even make the plot itself, it writes out a shell script to do
that, thus ``\texttt{pnuglot mydata}'' creates \texttt{mydata.gpl}
which then runs gnuplot, \LaTeX, and dvips to create
\texttt{mydata.eps}.  The advantage of this indirect method is that the
default gnuplot commands in \texttt{mydata.gpl} can subsequently be
edited to suit the user's taste.  Adding a label or choosing a different
line width is, for example, a pretty trivial matter.  Needless to say,
all labels are in \LaTeX\ and Type 1 fonts are selected to make the EPS
file nicely scalable.

\smallskip
\texttt{turnoff} switches off (and on) the evaluation of certain parts
of the amplitude, which is a handy thing for testing.  For example,
``\texttt{turnoff box}'' switches off all parts of the amplitude with
`box' in their name.  Invoking \texttt{turnoff} without any argument
restores all modules.


\section{Libraries and Makefiles}

The Fortran code is organized in \FC\ 4 into a main code directory, 
which contains the main program and all its prerequisite files, and 
subsidiary `folders' (subdirectories to the main code directory).  The 
default setup looks like this:
\begin{center}
\begin{picture}(200,100)
\Text(55,90)[b]{main code directory}
\Text(55,80)[b]{\small (created by \texttt{SetupCodeDir})}
\Line(0,77)(110,77)
\Line(55,77)(55,62)
\Line(55,62)(65,62)
\Text(70,62)[l]{\texttt{squared\uscore me/}}
\Text(70,52)[l]{\small (generated by \texttt{WriteSquaredME})}
\Line(45,77)(45,39)
\Line(45,39)(65,39)
\Text(70,39)[l]{\texttt{renconst/}\hphantom{p}}
\Text(70,29)[l]{\small (generated by \texttt{WriteRenConst})}
\Line(35,77)(35,16)
\Line(35,16)(65,16)
\Text(70,16)[l]{\texttt{util/}\hphantom{p}}
\Text(70,6)[l]{\small (comes with \FC)\hphantom{p}}
\end{picture}
\end{center}
Each folder is equipped with its own makefile which makes a library of 
the same name, \eg the makefile in \texttt{util/} makes the library 
\texttt{util.a}.

These sub-makefiles are orchestrated by the master makefile.  Libraries
required for the main program are listed in the \texttt{LIBS} variable
and built automatically by invoking the sub-makefiles:
\begin{verbatim}
  LIBS = squared_me.a renconst.a util.a
\end{verbatim}
The master makefile is no longer overwritten in the code-generation 
process and is treated like the other driver programs, \ie a customized
copy can be saved in the local drivers directory.

Occasionally it is useful to have more than one instance of
\texttt{squared\uscore me} (or \texttt{renconst}), \eg when computing an
hadronic cross-section to which several partonic processes contribute. 
Both \texttt{WriteSquaredME} and \texttt{WriteRenConst} have the
\texttt{Folder} option, with which a unique folder name can be chosen,
and the \texttt{SymbolPrefix} option, with which the symbols visible to
the linker can be prefixed with a unique identifier.  This identifier is
inserted only at compile time and can easily be changed in the
sub-makefile at any time.

The \texttt{util} library is a collection of ancillary routines which
currently includes: \\
-- System utilities (log file management), \\
-- Kinematic functions (Pair, Eps, $\dots$), \\
-- Diagonalization routines (Eigenvalues, $\dots$), \\
-- Univariate integrators (Gauss, Patterson), \\
-- Multivariate integrators (\cuba\ library). \\
Older versions of \FC\ used to include these directly into the main
program.  The library version has the advantage that the linker selects
only the routines that are actually needed, and furthermore it is
straightforward to add new code.  The \texttt{util.a} library is
compiled once when \FC\ is installed and then copied to the main code
directory, thus avoiding unnecessary compiles.


\section{Summary}

\FC\ 4 is the latest release of the Mathematica package \FC\ for
computing Feynman diagrams.  It has many new features enhancing in
particular performance, convenience, and modularity.

The package is available as open source and stands under the GNU library 
general public license (LGPL).  It can be obtained from the Web site
\texttt{http://www.feynarts.de/formcalc}.

Installation should be particularly painless with the new
\texttt{FeynInstall} script.  The user only has to answer `y' when the
script asks whether to install \FC.


\newcommand{\volyearpage}[3]{\textbf{#1} (#2) #3}
\newcommand{\cpc}{\textsl{Comp.\ Phys.\ Commun.} \volyearpage}
\newcommand{\npps}{\textsl{Nucl.\ Phys.\ Proc.\ Suppl.} \volyearpage}

\begin{flushleft}

\end{flushleft}

\end{document}